\documentclass[12pt]{iopart}
\usepackage{graphicx}
\usepackage{color}
\usepackage{epstopdf}
\DeclareGraphicsExtensions{.eps}

\begin{document}
\title[]{Density-induced geometric frustration of ultra-cold bosons in optical lattices}

\author{T. Mishra, S. Greschner \& L. Santos}

\address{$^1$ Institut f\"ur Theoretische Physik, Leibniz Universit\"at Hannover, 30167~Hannover, Germany}

\ead{tapan.mishra@itp.uni-hannover.de}

\begin{abstract}
A density-dependent gauge field may induce density-induced geometric frustration, leading to a non-trivial interplay between density modulation and frustration, which 
we illustrate for the particular case of ultra-cold bosons in zig-zag optical lattices with a density-dependent hopping. We show that the density-induced frustration leads in a rich landscape of quantum phases, including 
Mott and bond-order insulators, two-component superfluids, chiral superfluids, and partially-paired superfluids. We show as well that the density-dependent hopping results 
in an effective repulsive or attractive interactions, and that for the latter case the vacuum may be destabilized leading to a strong compressibility. Finally, we discuss how the predicted phases 
may be experimentally observed and characterized in time-of-flight measurements using their characteristic signatures in the momentum distribution.

\end{abstract}
\submitto{\NJP}
\maketitle

\section{Introduction}
Geometric frustration, especially in low-dimensional systems, results in the stabilization of unusual quantum phases~\cite{lewenstein}.
Ultra-cold atoms in optical lattices constitute an excellent environment for the study of the effects of geometric frustration in quantum many-body systems,
 due to the exquisite experimental control available over the lattice geometry, including triangular~\cite{Struck2012} and kagome~\cite{Jo2012} lattices, and even lattices 
 with variable geometry~\cite{Wirth2011,Tarruell2012}. Geometric frustration may result as well from the careful manipulation of the hopping rates. 
 Lattice shaking has been already employed to change selectively the sign of the hopping rate along some directions 
 in the lattice~\cite{Struck2011}, or even to obtain complex hopping rates~\cite{Struck2012}. Complex hopping rates have been realized as well 
 by means of Raman-assisted hopping, a technique that has recently allowed for the successful realization of synthetic magnetic fields~\cite{Aidelsburger2013, Miyake2013, Atala2014, Mancini2015, Stuhl2015}. 
  
Up to now, the created synthetic gauge fields, and in general the hopping rates in the lattice, are not affected by the atoms, i.e. the created gauge fields and lattice geometries are static. 
Recently, it has been proposed that laser-assisted hopping may result in the realization of density-dependent gauge fields~\cite{Keilmann2011,Greschner2015a,Greschner2015b,Bermudez2015}
for which the hopping rates (in particular their phase) is dynamically modified depending on the local lattice occupation. In one-dimensional lattices
density-dependent gauge fields result in the anyon-Hubbard model~\cite{Keilmann2011,Greschner2015a}, which is characterized by gauge-driven Mott-insulator to superfluid transitions, 
Mott phases at negative on-site interactions, and by the appearance of novel phases~(in particular the so-called partially-paired superfluid~\cite{Greschner2015a}). 
In higher-dimensions, laser-assisted hopping may induce under proper conditions density-dependent magnetism, which results in a non-trivial interplay 
between density modulation and chirality~\cite{Greschner2015b}.

\begin{figure}[t]
\begin{center}
\includegraphics[width=3.5in]{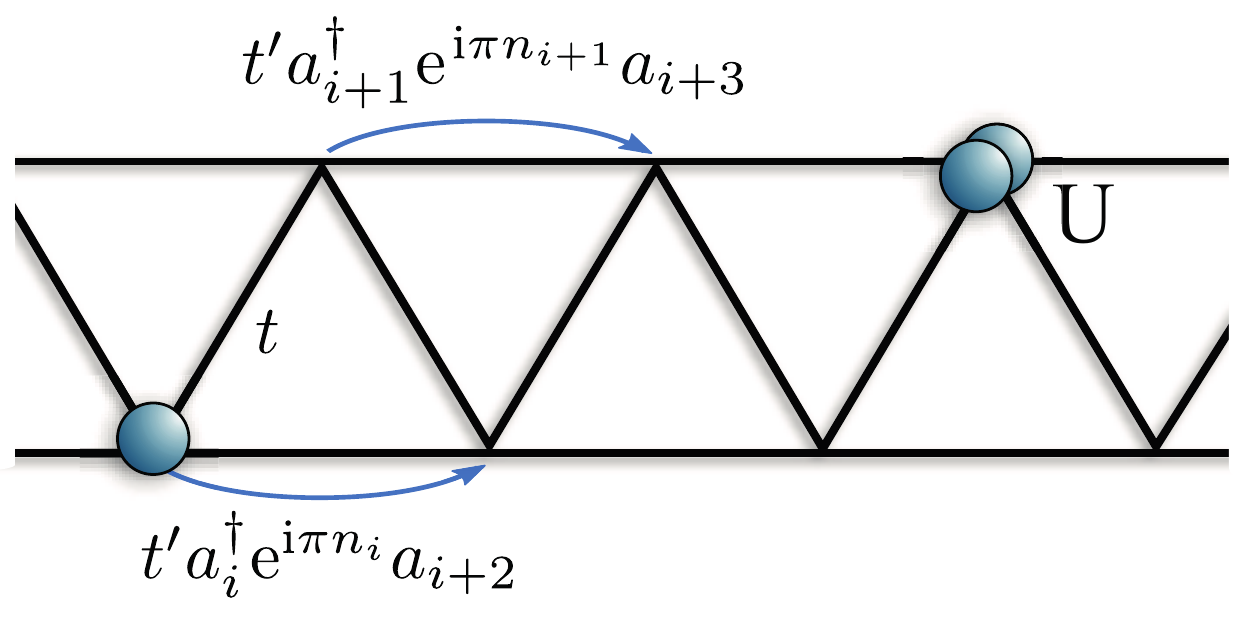}
\end{center}
\caption{(Color online) Schematic diagram of the zig-zag lattice with density-dependent next-to-nearest neighbor hopping considered in this paper.}
\label{fig:lattice}
\end{figure}

These recently proposed techniques for creating density-dependent hopping rates not only allow for the creation of density-dependent magnetism; more generally 
they open interesting possibilities for the study of lattice models with an occupation-dependent geometric frustration.
In this paper, we will illustrate the non-trivial interplay between lattice occupation and frustration in these models for the particular case of 
ultra-cold atoms in zig-zag lattices~(see Fig.~\ref{fig:lattice}). Zig-zag lattices are equivalent to one-dimensional lattices with nearest- and next-to-nearest-neighbor hoppings, 
and have been extensively investigated in the presence of frustration~\cite{Kolezhuk2002, Vekua2007, Hikihara2000, Hikihara2002, Hikihara2008,Hikihara2010,Greschner2013,Mishra2013,Mishra2014,Mishra2015}. 
They may be easily realized with ultra-cold atoms in optical lattices by superimposing incoherently a one-dimensional lattice and a triangular lattice, as shown in Ref.~\cite{Greschner2013}. 
We show below, that the occupation-dependent frustration results in a very rich landscape of insulator and superfluid phases, including chiral superfluids, two-component superfluids, 
Mott insulator phases with string order, bond-order insulators, and partially-paired superfluids. 

\section{Model and Method}
We consider a system of ultra-cold bosons in a zig-zag optical lattice with density dependent frustration as depicted in Fig.~\ref{fig:lattice}. The hopping rate along the 
rungs~(legs) is denoted as $t$~$(t')$. The coupling constant $U$ represents the two-body on-site repulsion between the atoms. Density-dependent frustration is realized by associating 
a density dependent Peierls phase to $t'$~(for a discussion on how these occupation dependent phases may be created by means of Raman-assisted hopping we refer to the detail discussions in 
Refs.~\cite{Greschner2015a,Greschner2015b}). The system is described by the Bose-Hubbard Hamiltonian:
\begin{equation}
\!\!\!\!\!\!\!\!\!\!\!H =- t \sum_{i} (a_{i}^{\dagger}a_{i+1}^{\phantom \dagger}+H.c.) -t'\sum_{i} (a_{i}^{\dagger}e^{i\pi n_i}a_{i+2}^{\phantom \dagger}+H.c.)+\frac{U}{2}\sum_i n_i(n_i-1)
 \label{eq:ham} 
\end{equation}
where $a_i^{\dagger}$, $a_i^{\phantom \dagger}$, and $n_i=a_i^{\dagger}a_i^{\phantom \dagger}$ are creation, annihilation, and number operators
for bosons at site $i$. As discussed in Refs.~\cite{Greschner2015a,Greschner2015b} employed to created the density-dependent Peierls phase results as well 
in an on-site two-body hardcore constraint~(2BHCC), $(a_i^\dagger)^3=0$, i.e. a maximum of two bosons may occupy a given site. 
For simplicity, we set in the following $t=1$ as the energy scale. 

Recent studies on ultra-cold bosons with 2BHCC in a fully-frustrated zig-zag lattice~(occupation-independent $t'<0$ and $t>0$) have 
shown that the system undergoes at unit filling and $U=0$ a transition from a superfluid~(SF) to a Haldane insulator~(HI) 
and then to a chiral superfluid~(CSF) phase as a function of frustration~(i.e. as a function of $|t'|/t$) 
~\cite{Greschner2013}. For $U<0$ the system exhibits transitions to other phases such as pair-superfluid~(PSF) and density wave~(DW). 
On the other hand a system of hardcore bosons ($U=\infty$) in a fully-frustrated zig-zag lattice resembles an isotropic $J_1-J_2$ model with some quantitative differences. 
The phase diagram of this system is dominated by the bond-order~(BO) phase at intermediate $|t'|$ values, whereas for large $t'$ the system exhibits a CSF~\cite{Mishra2013,Mishra2014,Mishra2015}. 

In this paper we are especially interested in the effects that the density-dependent sign of the next-to-nearest neighbor hopping in Model~(\ref{eq:ham})  
(and hence the corresponding occupation-dependent geometric frustration) has on the ground-state properties of the system. We will hence 
restrict for simplicity to the case $U=0$~(note however that the system remain effectively interacting due to the 2BHCC, and the occupation-dependent 
hopping itself), obtaining the ground-state phase diagram as a function of the frustrated hopping and the lattice filling using the
density matrix renormalization group~(DMRG) method~\cite{White1992,Schollwock2005} for up to $120$ lattice sites and $300$ density matrix eigenstates. 

\section{\bf Single particle dispersion}

Important aspects of the physics of the zig-zag lattice may be understood from its single particle dispersion, valid for the limit of low lattice filling, $\rho\to 0$
\begin{equation}
\epsilon(k)= - 2t \cos k -2 t' \cos 2k
\end{equation}
For $t'<0$ the model is frustrated, and for $|t'|/t>1/4$ the dispersion relation presents two degenerate minima at distinct momenta 
$k=\pm Q \equiv\pm \arccos(t/4|t'|)$. Due to this degeneracy the effect of interactions becomes crucial for selecting a particular ground-state.
Typically the two phases are found in the density-independent model~\cite{Kolezhuk2012}: a two-component (2SF) phase, in which the particles 
occupy both minima equally, and a CSF phase, with all particles quasi-condensing in one of the two minima, which is spontaneously selected.

Due to the 2BHCC, in the limit $\rho\to 2$, we may understand singly occupied sites (singlons) as moving 
on top of the fully occupied lattice $\cdot|2\rangle|2\rangle|2\rangle\cdot$. The singlons experience the effective dispersion:
\begin{equation}
\epsilon(k)= - 2 t \cos k + 2 t' \cos 2k
\end{equation}
Hence, a sufficiently large density induces frustration for $t'>0$, which is characterized as well by the presence of two degenerate minima in $\epsilon(k)$. 
We show below that this leads to the appearance of CSF and 2SF phases at large fillings.

\begin{figure}[t]
\begin{center}
\includegraphics[width=2.8in]{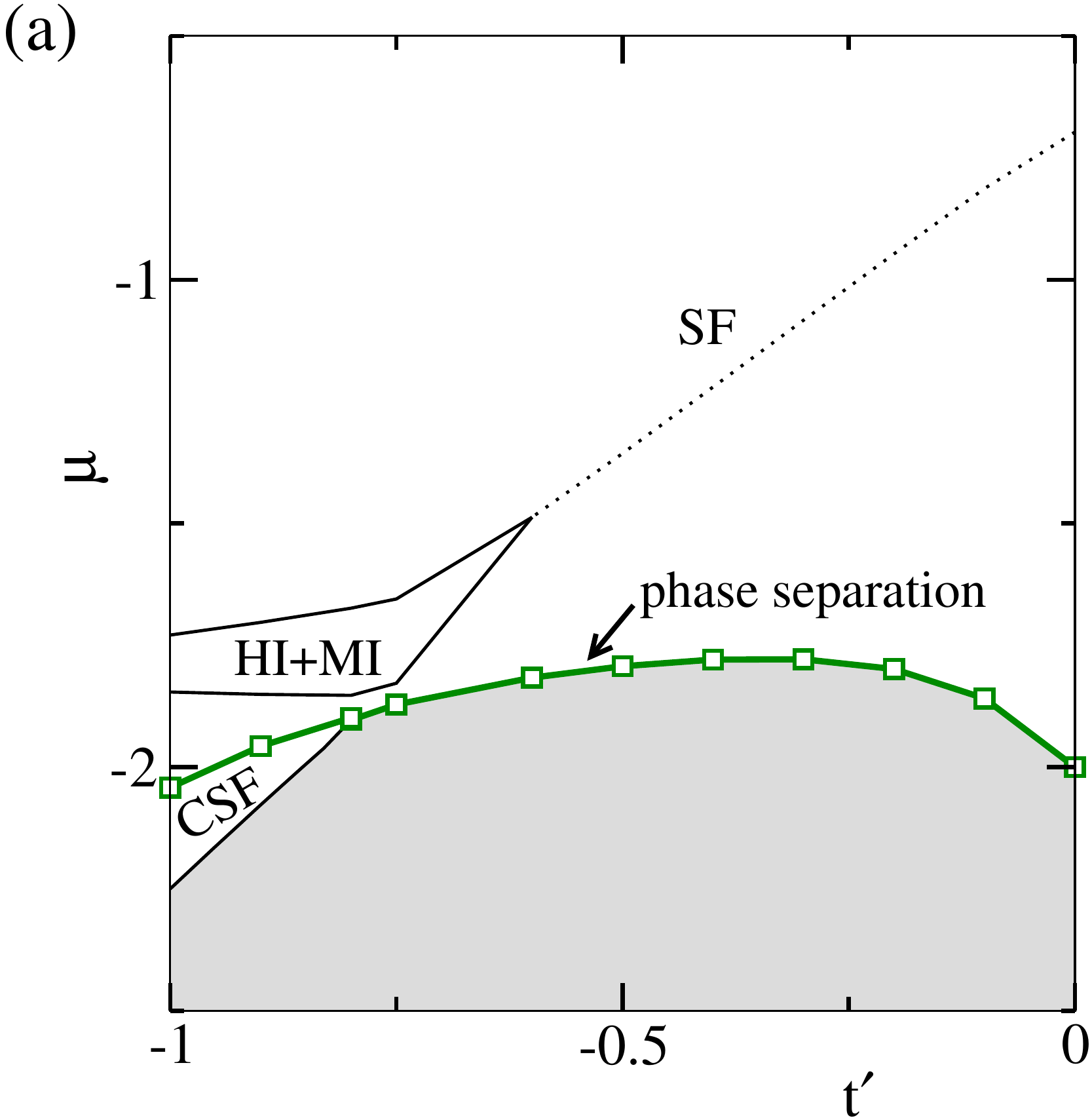}
\hspace{0.5cm}
\includegraphics[width=2.8in]{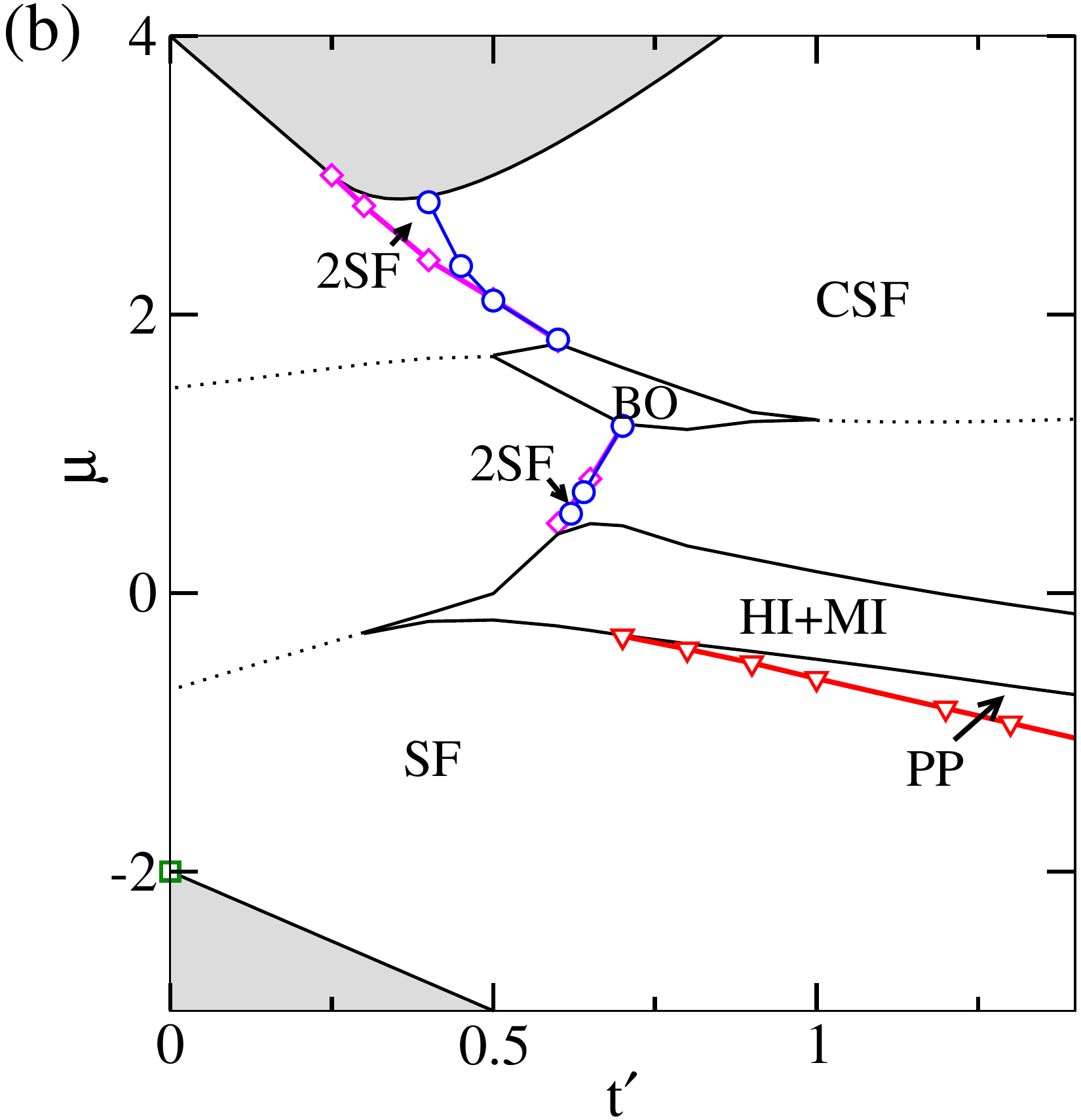}
\end{center}
\caption{(Color online) Ground-state phase diagram of Model~(\ref{eq:ham}) showing the phases for $t'<0$~(a) and $t'>0$~(b). The gapped MI+HI and BO phases are 
bounded by the black solid curves. For  $t>0'$ the PP phase is below the MI+HI phase, and the SF-PP transition is marked by red triangles. 
The 2SF phase is bounded between the magenta diamonds and blue circles. 
The CSF phase~(at the right of the blue circles) covers a broad range of the phase diagram. For $t<0'$, the abrupt density jump~(phase separation) is marked by green squares. In both diagrams, the grey regions represent the empty~($\rho=0$) and full~($\rho=2$) states.}
\label{fig:phasedia}
\end{figure}

\begin{figure}[t]
\begin{center}
\includegraphics[width=5.in]{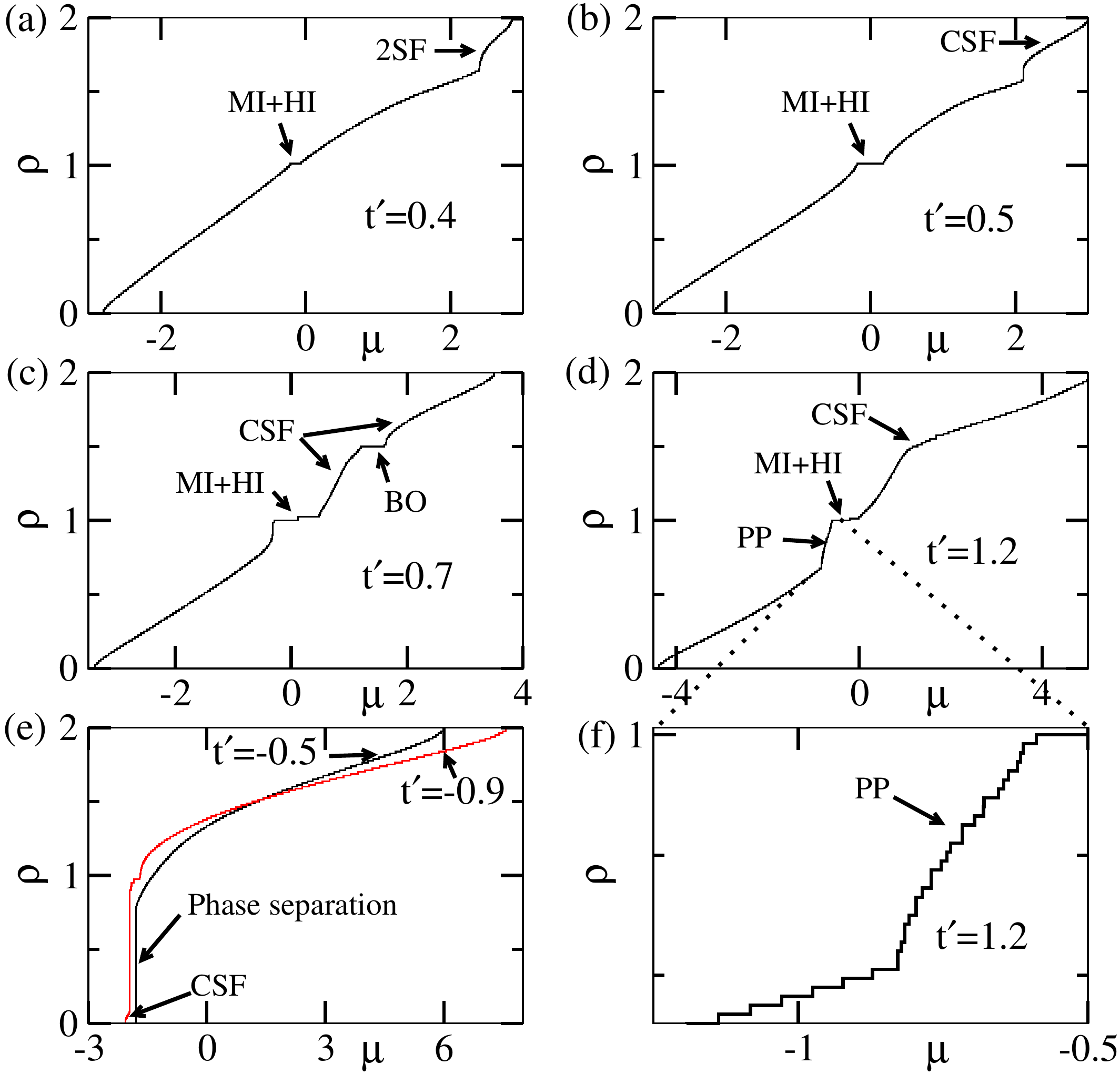}
\end{center}
\caption{(Color online) Curves $\rho(\mu)$ for $t'=0.4$~(a),$t'=0.5$~(b), $t'=0.7$~(c), $t'$=1.2~(d), 
and  $t'=-0.5$ and $-0.9$~(e). Figure (f) shows an enlarged view of the PP region in Fig. (d).}
\label{fig:rhomu}
\end{figure}


\section{Density-induced frustration}

We consider in this section first the case of $t'>0$, i.e. a model which is unfrustrated for low filling but becomes frustrated at large filling. 
In the next section we briefly discuss the case of $t'<0$. 
Figure ~\ref{fig:phasedia}(b) summarizes our DMRG results for the ground-state phases of Model~(\ref{eq:ham}) with $U=0$. 

\subsection{Gapped phases}

As mentioned above, a fully frustrated zig-zag lattice with 2BHCC exhibits for $U=0$ at unit filling, and solely due to frustration, a gapped HI between the gapless SF and CSF phases. 
Since in Model (\ref{eq:ham}) frustration emerges with growing filling we may expect also in this model 
the appearance of gapped phases for $U=0$. In order to distinguish between gapped and gapless phases we evaluate the chemical potential
$\mu$ obtained from the minimization of $E(L,N)-\mu N$, where  $E(L,N)$ is the ground-state energy for $N$ bosons in $L$ sites~\cite{Hikihara2008}.
Figure~\ref{fig:rhomu} depicts the chemical potential as a function of the lattice filling for different values of $t'$. Plateaus in this 
graph constitute a clear signature of the existence of gapped phases. Although the lattice plaquettes are frustrated for a sufficiently large filling, 
for small $t'$ the system remains in the SF phase for all densities.  As the hopping strength increases a gapped phase appears for $\rho=1$ for $t'>0.3$, whereas a second plateau 
is observed for $\rho=3/2$ for $t'>0.6$. These plateaus appear for a range of 
intermediate values of $t'$~(see Figs.~\ref{fig:rhomu}(a)--(d)) and then vanish for large $t'$. We plot the extrapolated values of the chemical
potentials corresponding to these plateaus in Fig.~\ref{fig:phasedia}~(lobes bounded by black curves). 

\begin{figure}[t]
\begin{center}
\includegraphics[width=5.in]{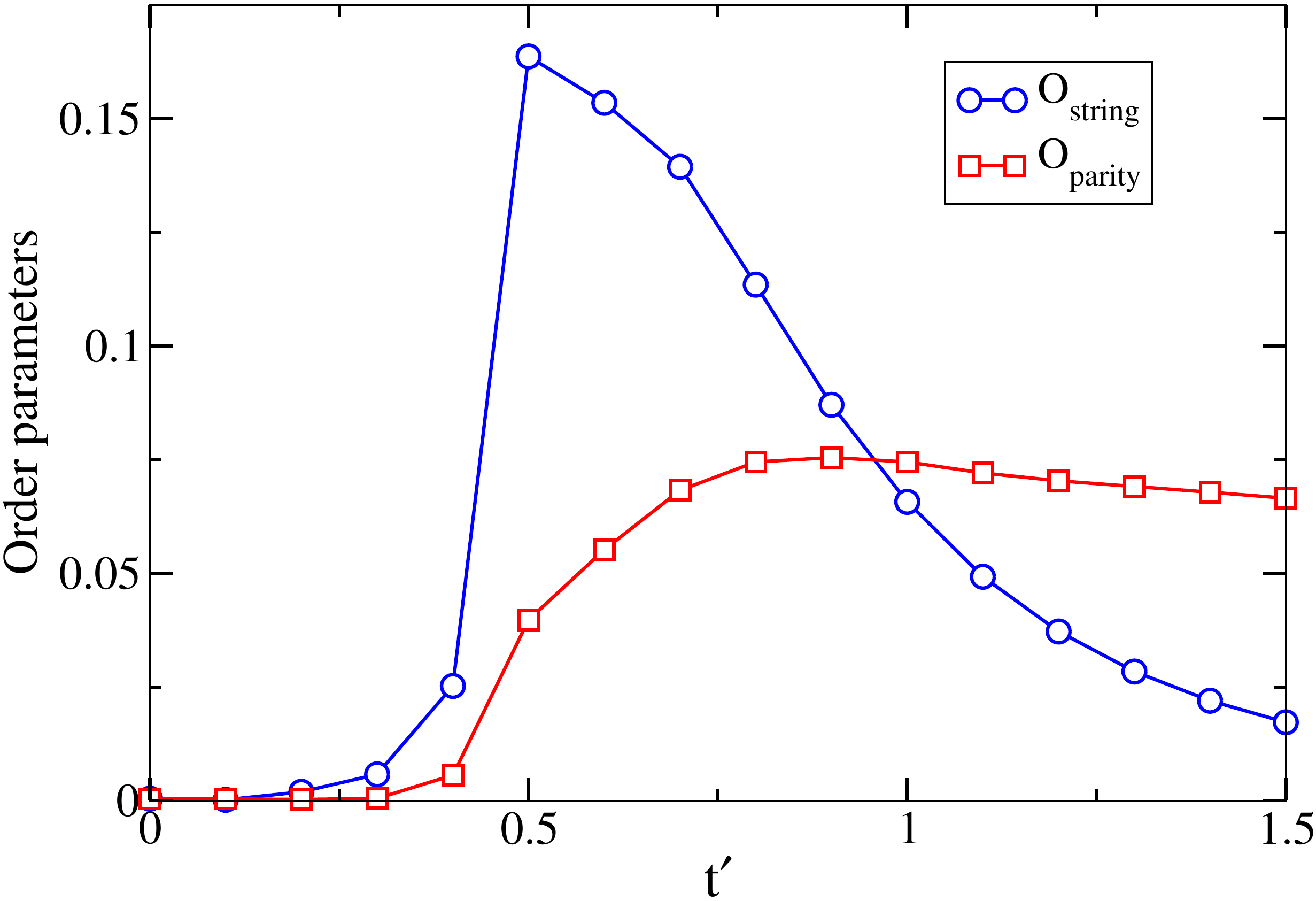}
\end{center}
\caption{(Color online) Extrapolated values of the string and the parity order parameters with respect to $t'$ for $\rho=1$. 
The simultaneous presence of both orders characterizes the MI+HI phase.}
\label{fig:strparity}
\end{figure}

The gapped phase at $\rho=1$ is a MI resulting from the density-induced frustration. The MI phase is characterized by a finite parity-order parameter
\begin{equation}
 O_{parity}=\lim_{|i-j|\rightarrow \infty} \langle (-1)^{\sum_{i<l<j} \delta n_l}\rangle,
 \label{parity}
\end{equation}
with $\delta n_j=1-n_j$, which has been recently measured experimentally using single-site resolution~\cite{kuhr}. 
In addition, and contrary to the usual MI, the gapped phase at $\rho=1$ presents a non-zero string-order parameter
\begin{equation}
 O_{string}=\lim_{|i-j|\rightarrow \infty}\langle\delta n_i(-1)^{\sum_{i<l<j} \delta n_l}\delta n_j\rangle,
 \label{string}
\end{equation}
which is typically the signature of a HI. Figure~\ref{fig:strparity} shows that both order parameters become finite 
around $t'\simeq 0.3$ where the gap opens. We denote this phase as MI+HI phase in the phase diagram of 
Fig.~\ref{fig:phasedia}.

\begin{figure}[t]
\begin{center}
\includegraphics[width=5.in]{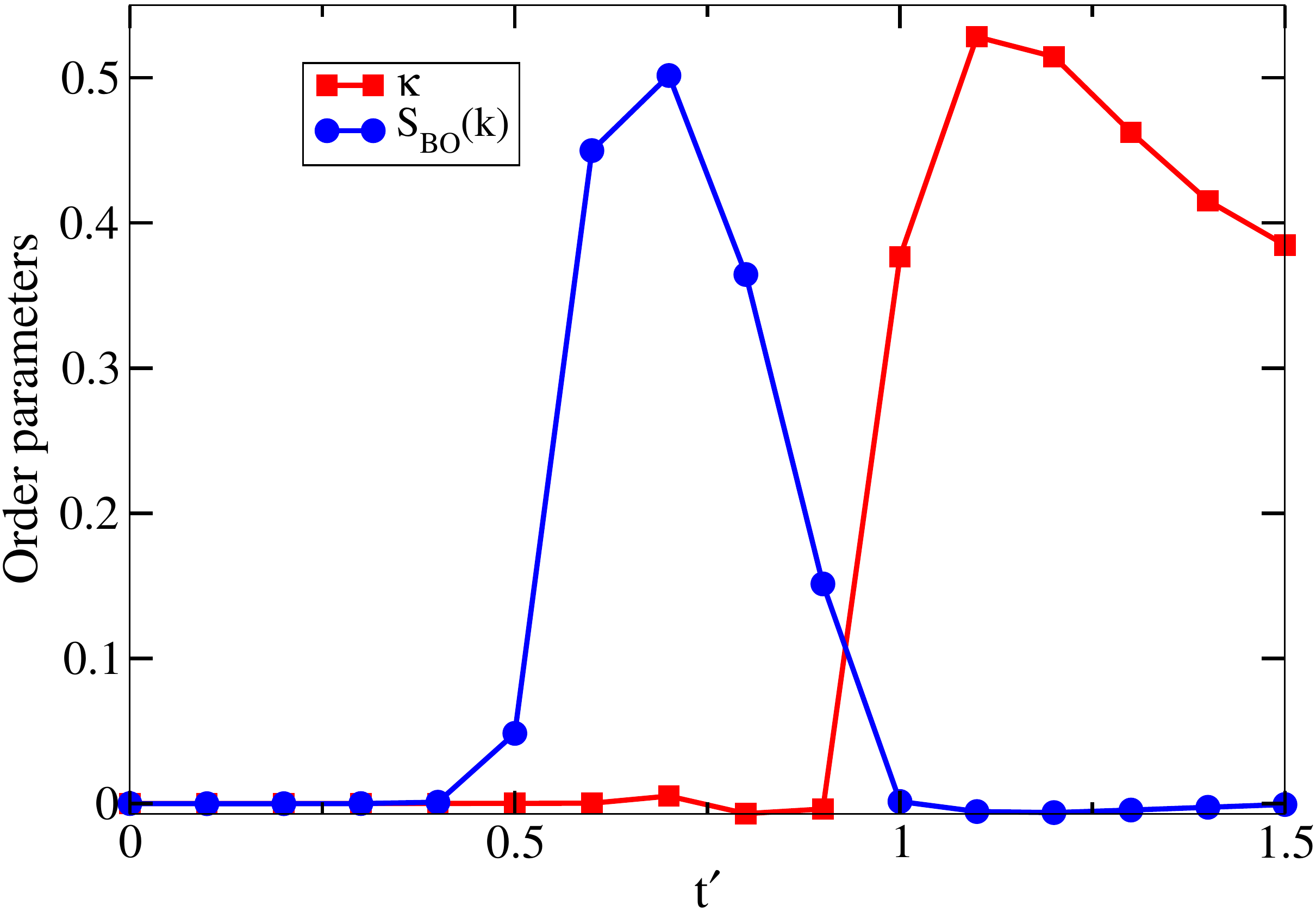}
\end{center}
\caption{(Color online)  Extrapolated values of the bond-order structure factor $S_{BO}(k=\pi)$ and the chiral order parameter $\kappa$, in the SF-BO-CSF region.}
\label{fig:bostr}
\end{figure}

\begin{figure}[t]
\begin{center}
\includegraphics[width=5.in]{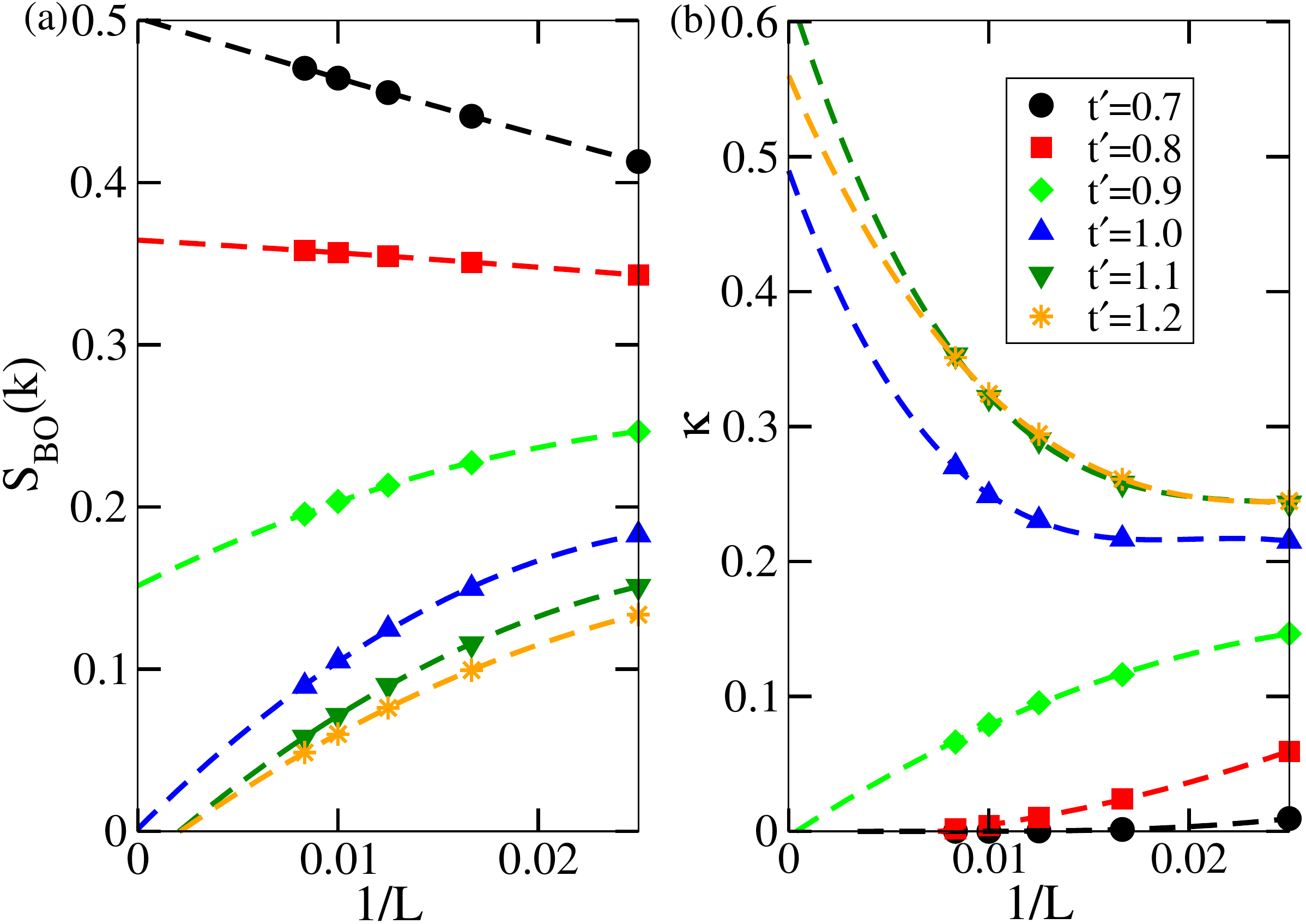}
\end{center}
\caption{(Color online)  Finite size scaling of $S_{BO}(k=\pi)$~(a) and $\kappa$~(b) across the BO-CSF boundary. }
\label{fig:bostrther}
\end{figure}

In contrast, the gapped region at $\rho=3/2$ is a BO phase that results as well solely from the density-induced geometric frustration. 
Due to the 2BHCC, doublons (i.e. doubly-occupied sites) on top of the MI phase with all the sites occupied by one particle may be considered as hard-core bosons.
The full state ($\rho=2$) may be hence understood as an insulator with unit filling occupation ($\rho_D=1$) of these hard-core bosons.
The appearance of the BO phase can be hence understood from what is known of hard-core bosons in zig-zag lattices~\cite{Mishra2013,Mishra2014,Mishra2015}. 
At density $\rho=3/2$~($\rho_D=1/2$) the doublons minimize the energy arrange'ng themselves along the rungs of the ladder in a periodic pattern, that results in bond order. 
The existence of bond ordering is confirmed by a finite peak in the BO structure factor 
\begin{equation}
 S_{BO}(k)=\frac{1}{L^2}\sum_{i,j}{e^{ik(i-j)}\langle B_i B_j\rangle},
\label{eq:str}
\end{equation}
where $B_i=a_i^\dagger a_{i+1}^{\phantom \dagger}+a_{i+1}^\dagger a_i^{\phantom \dagger}$. 
By performing a finite size extrapolation of $S_{BO}(k=\pi)$ using system sizes of $L=40, 60, 80, 100, 120$ we establish the existence of a BO phase for $0.4~< t' ~<1.0$ as shown in 
Fig.~\ref{fig:bostr}~(blue circles). Upon increasing $t'$ the system enters into the CSF phase, similarly as for the case of hardcore bosons~\cite{Mishra2015}. 
The transition to the CSF phase at $\rho=1.5$ is obtained by plotting the chiral order parameter 
\begin{equation}
 \kappa=\lim_{\left|i-j\right|\gg1} \left<\chi_i\chi_j\right>
 \label{eq:chi}
\end{equation}
where $\chi_i=i(a_i^\dagger a_{i+1}^{\phantom \dagger} -a_{i+1}^\dagger a_i^{\phantom \dagger})$, is the current-current correlation 
function. In Fig.~\ref{fig:bostr}~(red squares)
we plot the extrapolated values of $\kappa$ using system sizes of $L=40, 60, 80, 100, 120$ which shows a transition to the CSF phase at 
$t'\sim 1.0$. 
The finite size scaling of $S_{BO}(\pi)$ and $\kappa$ is shown in Fig.~\ref{fig:bostrther}. It can be clearly seen that $S_{BO}(k)$ slowly goes 
to zero as $t'$ increases and at the same time the value of $\kappa$ becomes finite at $t'\simeq 1.0$.

\subsection{Gapless phases}

Whereas for low fillings the system is a simple SF phase, for $t' > 0.7$ and below $\rho=1$ a different phase appears, characterized by an increase of the filling 
for growing $\mu$ in an unordered sequence of steps of 
one and two bosons~(see Fig.~\ref{fig:rhomu}(f)). This feature is a signature of the so-called partially-paired(PP) phase,  a 
two-component superfluid phase of doublon- and holon-dimers characteristic of the one-dimensional anyon-Hubbard model~\cite{Greschner2015a}. 
In Model~(\ref{eq:ham}) the PP phase results from the quasi-anyonic character of the chains built up of even and odd sites, which become fully decoupled only in the limit $t'\gg t$. 
The PP phase survives in the region below $\rho=1$ MI+HI phase for large values of $t'$ as shown in the Fig.~\ref{fig:phasedia}. 
As we will show below the PP-phase exhibits a multi-peak momentum distribution.
 
\begin{figure}[t]
\begin{center}
\includegraphics[width=5.in]{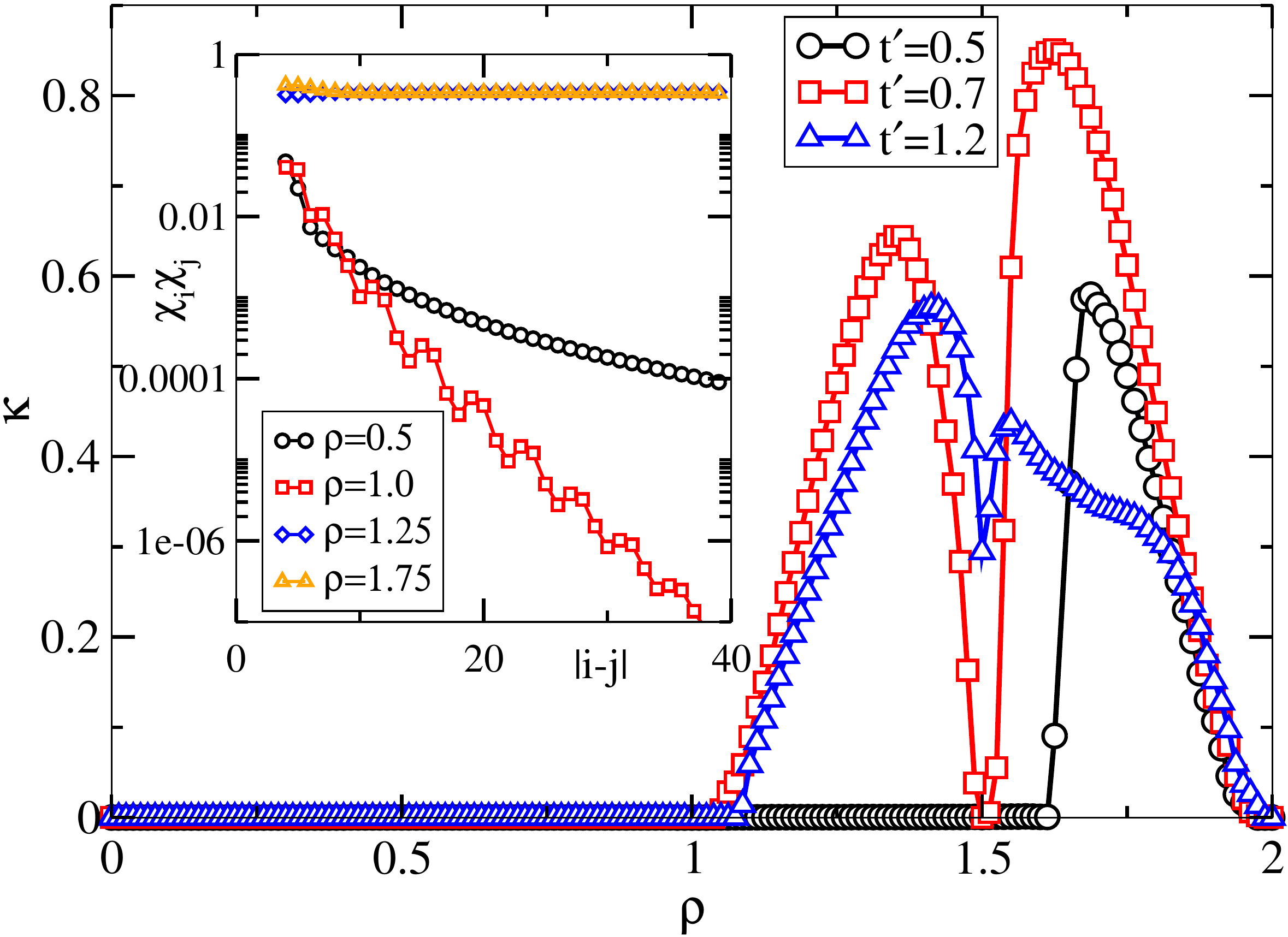}
\end{center}
\caption{(Color online)  Chiral order parameter $\kappa$ with respect to $\rho$ for different $t'$ at the transition to the CSF phase~(see text). The inset shows the current-current correlation 
$\chi_i\chi_j$ for different fillings.}
\label{fig:chiral}
\end{figure}

For $\rho>1$, the system is a CSF, which is also stable for 
$\rho=3/2$ when $t' > 1.0$ as discussed above. The CSF phase extends up to $\rho=2$. In Fig.~\ref{fig:chiral} we plot 
the chiral order parameter $\kappa$ as a function of $\rho$ for three representative values of $t'$. 
For $t'=0.5$~(black circles) $\kappa$ goes from zero to a finite value marking the SF~to~CSF phase transition.
For $t'=0.7$~(red squares), the region of finite $\kappa$ is split into two parts by a region of $\kappa=0$, marking the CSF-BO-CSF transitions~(see Fig.~\ref{fig:phasedia}(b)).
For $t'=1.2$~(blue triangles) the intermediate BO region has disappeared.
The current-current correlation function $\chi_i\chi_j$ is plotted in different regions of the phase diagram in the inset of Fig.~\ref{fig:chiral}. There is a sharp decay of the correlation function when the system is 
in the SF phase, whereas in the CSF phase the value of $\chi_i\chi_j$ shows a long-range correlation. 

Apart from all these phases we also see signatures of a 2SF phase in the vicinity of the gapped phases. The transition to the 2SF phase is marked by a kink in the $\rho(\mu)$ 
curve~(see Fig.~\ref{fig:rhomu}(a)). We also see similar feature in the region between the MI+HI phase and the BO phase as 
shown in Fig.~\ref{fig:phasedia}(b). 

\section{Frustration decreasing with the density}

For $t'<0$ the system is frustrated at $\rho\to 0$, and for growing density the frustration is destroyed. One could hence naively expect an approximate inversion of the 
graph obtained for $t'>0$. The situation is however rather different. For all $t'<0$ we observe a large 
density jump for a fixed chemical potential~(see Fig.~\ref{fig:rhomu}(e) and the green squares in Fig.~\ref{fig:phasedia}(a)).  
For $|t'|<0.8$ the density jump starts from the vacuum, $\rho=0$, whereas for $|t'|>0.8$ there exists a finite CSF region at low densities. 
At $\rho=1$ and $|t'|>0.6$ we observe the MI+HI phase. Further increase in the density results in the SF phase. 
The presence of the strong density jump prevents the appearance of a BO phase at $\rho=0.5$, the 2SF phase and the PP phase.

The presence of the density jump may be understood from the following simple argument. Due to the 2BHCC, $e^{i\pi n}=1-2n$. Although, the 
correction of the hopping depends on the number operator, we may understand the main effects of this correction by approximating $t'e^{i\pi n}\simeq t'_{eff}(\rho)=t'(1-\gamma\rho)$, where 
$\gamma>0$ is a proportionality constant. For $t'>0$, the single particle dispersion presents a single energy minimum at $k=0$, with density-dependent 
energy $E_0(\rho)=-2-2t'_{eff}(\rho)=E_0(0)+2t' \gamma\rho$. Note that the vacuum boundary is given by the curve $\mu=E_0(0)$. Since the energy increases with $\rho$ the vacuum boundary 
is stable, and we expect a continuous growing of $\rho$ as a function of $\mu$. 
For $t'<0$ the situation radically changes, as it is possible to see from the simple model above for $|t'|<1/4$. 
There is still a single minimum at $k=0$, but now with energy $E_0(\rho)=-2+2|t'_{eff}(\rho)|=E_0(0)-2|t'|\gamma\rho$, i.e. now the energy 
decreases with $\rho$. As a result, when $\mu=E_0(0)$, the density-dependent hopping pushes $E_0(\rho)$ even further below $\mu$, and hence $\rho$ experiences a 
large jump (only arrested by 
corrections to the simple toy model).

\section{Effective interaction in the limit $\rho\to0$ and $\rho\to 2$}

In the limits $\rho\to0$ and $\rho\to 2$ we may derive an description of the system properties by means of a two-particle scattering problem~\cite{Kolezhuk2012, Azimi2014}. 
A general two-particle state may be described by
\begin{equation}
\left| \Psi_K \right> = \sum_x c_{x,x} \left(b_x^\dagger\right)^2 \left|0\right> + \sum_{x,y>x} c_{x,y} b_x^\dagger b_y^\dagger \left|0\right>.
\end{equation}
Due to the conservation of total momentum  $K=k_1+k_2$ in the scattering process one can express the amplitudes as $c_{x,x+r} = C_r \e^{i K (x+\frac{r}{2})}$.  
The Schr\"odinger equation $H \left|\Psi\right> = \Omega \left|\Psi\right>$ for the two-particle problem leads to the following system of coupled equations  for the amplitudes $C_r$:
\begin{eqnarray}
&&\!\!\!\!\!\!\!\!\!\!(\Omega - 2 U) C_0 = -2 \sqrt{2} \left( t \cos\left(\frac{K}{2}\right) C_1 - {\rm i} t' \sin(K) C_2\right) \label{eq:dilute_2particle_eqs-1}\\
&&\!\!\!\!\!\!\!\!\!\!\Omega C_1 = -2  t \cos\left(\frac{K}{2}\right) \left( \sqrt{2} C_0 + s C_2\right) - 2 t' s \cos K \left(C_3 + C_1\right) \label{eq:dilute_2particle_eqs-2} \\
&&\!\!\!\!\!\!\!\!\!\!\Omega C_2 = -2  t \cos\left(\frac{K}{2}\right) \left( C_1 + C_3 \right) - 2 t'\left( \sqrt{2} {\rm i} \sin(K) C_0 + s \cos(K) C_4\right) \label{eq:dilute_2particle_eqs-3}\\
&&\!\!\!\!\!\!\!\!\!\!\Omega C_r = -2  t s \cos\left(\frac{K}{2}\right) \left( C_{r-1} + C_{r+1}\right) - 2 t' s \cos(K) \left( C_{r-2} + C_{r+2}\right)\, ,r\geq 3 \label{eq:dilute_2particle_eqs-4}
\end{eqnarray}
For $\rho\to 0$~($2$) we set $s=1$~($2$).
The density-dependent term in Model~(\ref{eq:ham}) just enters through the sine function in the first and third equation (for the usual Bose-Hubbard model it would be a cosine).
In the thermodynamic limit the energy is given by 
$\Omega = \epsilon(k_1) + \epsilon(k_2) = - 4 (J \cos(k) \cos(\frac{K}{2}) + J' \cos(2k) \cos(K))$
with the half relative momentum  $k=(k_1-k_2)/2$.
For the scattering of two particles in the vicinity of the minimum at $Q=0$ with momentum $k_1=Q+k$ and $k_1=Q-k$, i.e. total momentum $K=Q$, 
one may solve the system of equations with an Ansatz
\begin{equation}
C_{r}=\cos(kr +\delta) +v \e^{-\kappa_0 r}
\end{equation}
The coefficients $C_0$, $\delta$ and $v$ are determined by Eqs.~(9, 10, 11), 
hence, are affected by the density-dependent hopping.
The scattering lengths may be extracted from the scattering phase shift $\delta$, 
$a=\lim_{k\to 0} \cot(\delta)/k$.
One can relate the 1D scattering length to the amplitude of the contact interaction potential of the two-component Bose gas of mass $m$ as $g=-2/ma$. 
For $|g|m\ll 1$ we may employ a continuum Lieb-Liniger model~\cite{Kolezhuk2012}. In this limit, for $g<0$, the bosons
form an attractively interacting model, characterized by the presence of bound states or a collapsing wave function.

For $|t'|\ll 1$, and after some algebra, we obtain:
\begin{eqnarray}
\!\!\!\!\!\!\!\!\!\! g_{\rho \to 0} &=& \frac{U + 2 t' (2+2 t'+U)}{(1+t' (2+2 t'+U)}\simeq U + (4 - U^2) t'  , \\
\!\!\!\!\!\!\!\!\!\! g_{\rho \to 2} &=& \frac{2 (3 + U + t' (11 + 8 t' + 2 U))}{1 + 2 t' (4 + 4 t' + U)}\simeq 6 + 2 U - 2 t' (13 + 2 U (6 + U))
\end{eqnarray}
For $U=0$, $g_{\rho \to 0}=4t'$, and hence for $t'<0$ an effective attraction is realized, providing an alternative intuitive picture 
of the instability observed in the $\rho(\mu)$ curve. In contrast, $g_{\rho \to 2}=6 (1-6t')$ remains positive for any small $t'$, leading to 
an effective repulsive interaction even for $U\to 0$ due to the density-dependent frustration.

For the case of two degenerate minima these arguments may be generalized. One may consider both dispersion minima as two 
distinct species of particles. The low-energy properties are determined by both the scattering from particles of the same minimum 
and scattering of particles from different dispersion minima, which give rise to effective intra- and inter-species interactions. 
For the case of a dominant intra-species interaction both minima are occupied and a 2SF-phase is stabilized. 
A dominant inter-species coupling will lead to a CSF phase in the dilute limit. For a detailed calculation we refer to Ref.~\cite{Kolezhuk2012}.
For this case we determine the 2SF-CSF transition for $t' \approx 0.97 t$ for the case $\rho\to 2$ which only coincides 
qualitatively with our DMRG calculation(as shown in Fig.\ref{fig:phasedia} we 
could determine the boundary around $t' \approx 0.4 t$) showing that at large $t'/t$ the CSF phase is favored. 
This discrepancy has been reported already in Ref.\cite{Mishra2015,Kolezhuk2012,Mishra2015b} and is 
possibly due to the difficulty in determining the CSF-2SF boundary accurately at low or large fillings numerically because of vanishing chiral order parameter.

\section{Experimental signature}
\begin{figure}[!h]
\begin{center}
\includegraphics[width=6.in]{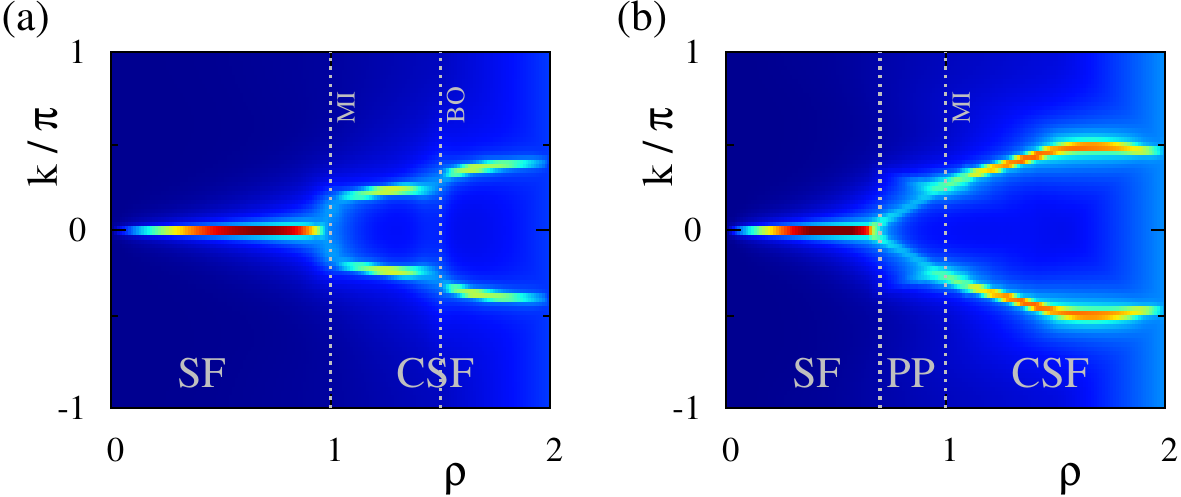}
\end{center}
\caption{(Color online) Momentum distribution function $n(k)$ for $t'=0.7$~(a) and $t'=1.5$~(b). Note the different form of $n(k)$ for the different phases of the system.}
\label{fig:mom}
\end{figure}

In order to obtain the signatures of the quantum phases we compute the momentum distribution function
\begin{equation}
n(k)=\frac{1}{L}\sum_{i,j}{e^{ik(i-j)}\langle{a^{\dagger}_{i}a_{j}}\rangle}.
\label{eq:mom}
\end{equation}
where $\langle{a^{\dagger}_{i}a_{j}\rangle}$ is the single particle correlation function. 
We plot $n(k)$ as a function of $\rho$ along two representative cuts through the phase diagram, $t'=0.7$ and $1.5$, as shown in Fig.~\ref{fig:mom}. 

For $t'=0.7$, as the system goes from SF~-~MI~-~CSF~-~BO~-~SCF as a function of $\rho$, 
we observe for growing $\rho$ regions of different momentum distributions: 
a single peak of $n(k)$ at $k=0$~(SF), no peak~(HI+MI at $\rho=1$), two peaks at $\pm k$~(CSF), no peak~(BO at $\rho=3/2$), and 
again two peaks at $\pm k$~(CSF).
For $t'=1.5$, we observe a markedly different dependence for growing $\rho$: a region with a single momentum peak~(SF), then a region with multiple peaks~(PP), 
no peak~(MI+HI at $\rho=1$), and two peaks at $\pm k$~(CSF).
The momentum distribution, which may be extracted from standard  time-of-flight measurements, constitutes hence an excellent experimental observable for the observation and 
characterization of the different phases.

\section{Conclusions}
Laser-assisted hopping may result in density-dependent magnetism, and more generally in density-dependent frustration. In this paper we have illustrated the consequences 
that density-induced frustration (together with 2BHCC) may have on the properties of a bosonic gas, for the particular case of bosons in a zig-zag lattice in which the 
sign of the next-to-nearest hopping (and hence the geometric frustration) depends on the local occupation. We have shown that, even in the absence of two-body interactions ($U=0$) 
the density-dependent frustration leads to a wealth of gapped~(MI+HI, BO) and gapless~(SF,2SF,CSF,PP) phases, 
since for $t'>0$ the system basically interpolates for growing filling from a non-frustrated non-interacting system to a fully frustrated system with an effective two-body repulsion induced by the 
density-dependent hopping. In contrast, for $t'<0$, frustration decreases with density, but the frustrated low density regime is unreachable due to the vacuum destabilization resulting from 
the effective inter-particle attraction induced for $t'<0$ by the density-dependent hopping. 
Finally, we have shown that the discussed phases may be experimentally revealed in time-of-flight measurements due to their different signature in the momentum distribution.

\ack
We acknowledge support from the Center for Quantum Engineering and Space Time Research and the Deutsche Forschungsgemeinschaft(Research Training Group 1729). 
Simulations were performed on the cluster system of the Leibniz Universit\"at Hannover.

\end{document}